# Fifth ASTROD Symposium and Outlook of Direct Gravitational-Wave Detection

From July 11 to July 13, 2012, Raman Research Institute (Bangalore, India) hosted the *Fifth International ASTROD Symposium on Laser Astrodynamics, Space Test of Relativity and Gravitational-Wave Astronomy.* Sixty-one persons attended the Symposium including 24 invited speakers, 15 professionals from various fields and 20 students (Fig. 1). The aim of this series of Symposia is to focus on various disciplines related to fundamental physics in space, to foster dialogues and to plan for the future. Previous ASTROD Symposia were held during September 21-23, 2001 (Beijing), June 2-3, 2005 (Bremen), July 14-16, 2006 (Beijing) and July 16-17, 2010 (Bremen).

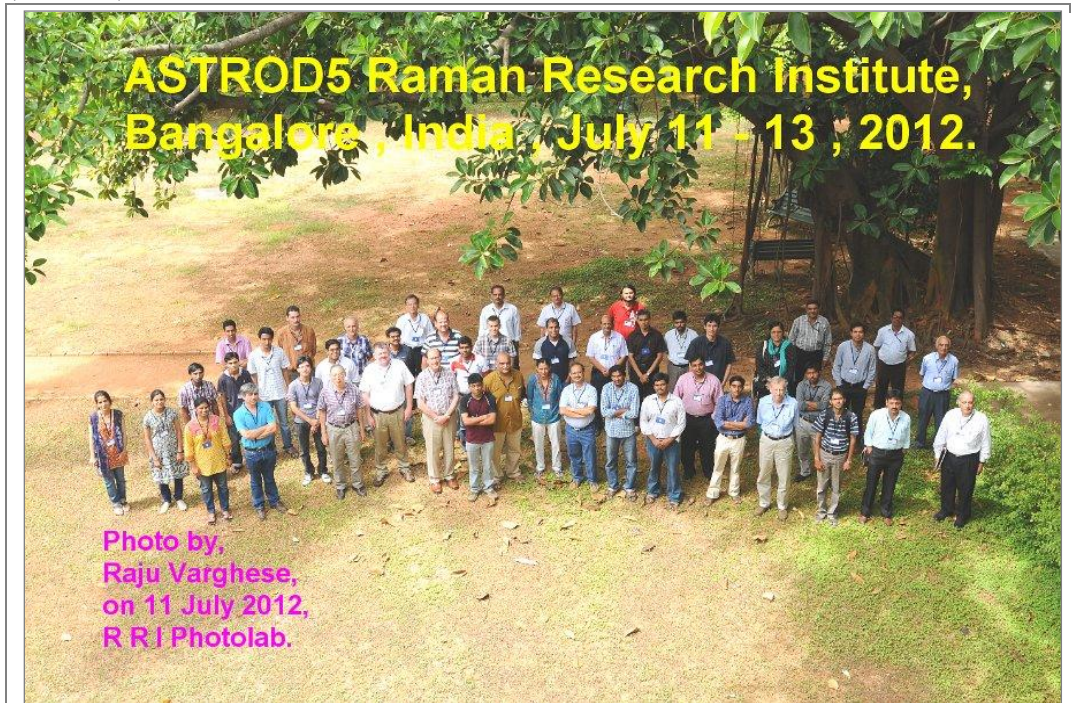

Fig. 1. ASTROD5 group photo

The inaugural address was made by P. Sreekumar of the Indian Space Research Organization (ISRO). The focus in the Symposium was on Gravitational Wave (GW) detection, *especially with space-based interferometers*. However, in view of the exciting proposals over the last couple of years in establishing a global network crucial for GW astronomy, the meeting began with ground based detectors. Daniel Sigg from the LIGO gave the status report on Ground-Based GW Interferometers. With the ongoing construction of Advanced LIGO (4 km arm length, in Hanford and Livingston, USA), Advanced Virgo (3 km arm length, in Cascina, Italy) and KAGRA (3 km arm length, in Kamioka, Japan), the first GW detection is anticipated around 2016 and to open an era of experimental GW Astronomy.

C. S. Unnikrishnan from TIFR summarised the initiatives by the IndIGO (Indian Initiative in Gravitational Wave Observations) Consortium during the past two years which has now materialized into concrete plans and project opportunities for instrumentation and research based on advanced interferometer detectors. The proposed LIGO-India project is a culmination of a 2 year long intense effort by IndIGO, which will foster integrated development of frontier gravitational wave research in India and will facilitate direct participation in global GW research with substantial contributions to gravitational wave astronomy. In the wake of these developments a surge of activity in precision metrology, instrumentation, data handling and computation are expected with LIGO-India as inspiration.

The laser-interferometric ground detectors are most sensitive to the high frequency band (10 Hz – 10 kHz). The major sources are coalescences of compact binaries of neutron stars (NSs) and stellar mass Black Holes (BHs), and core collapse supernovae. Other promising frequency bands for direct detection are the middle frequency band (0.1 Hz – 10 Hz) and the low frequency band (100 nHz – 0.1 Hz) utilizing space interferometric detectors, and the very low frequency band (300 pHz – 100 nHz) using Pulsar Timing Arrays (PTAs). The complete frequency classification of GWs is given in http://astrod.wikispaces.com/file/view/GW-classification.pdf.

Pulsars are ultra-stable clocks through precision timing. When very low frequency GWs pass by the line of sight of pulsars, they encode periodic signals on the arrival times of pulses and therefore an array of pulsars emitting pulses with ultra-stable periods can serve as a GW detector. The ground pulse receiver can be a single detector or multiple detectors or an array. Dick Manchester (CSIRO) and Bhal Chandra Joshi (NCRA) reviewed pulse timing and reported on the sensitivities of current pulsar timing arrays. With the current ranges of prediction of GW backgrounds from SMBH (supermassive black hole) - SMBH mergers, these background GWs would be detectable by PTAs (Pulsar Timing Arrays) around 2020. The present upper limits on the background from PTAs reach $10^{-15}$ level of the characteristic strain in the very low frequency band and already rule out models in which giant elliptical galaxies grow by merger alone.

*Direct detection of GWs with high signal to noise ratios would need space missions to realize.* Space GW detection is the main focus of ASTROD5 Symposium. In Bernard Schutz's (AEI) talk, GW sources for space detectors were reviewed extensively. These sources include: (i) Massive BHs ($10^3$--$10^8$ $M_{Sun}$): mass function, spin evolution as function of redshift z, sampling central black holes in ordinary galaxies, search for intermediate mass BHs (IMBH); (ii) Evolution of the Cosmic Web at high redshift: observation of objects before re-ionisation (BH mergers at $z \gg 10$), testing models of how massive BHs formed and evolved from seeds; (iii) Compact WD binaries in the Galaxy: catalogue of white-dwarf (WD) binary systems in the Galaxy, comparison with GAIA, precise masses and distances for many WD/NS/BH binaries. The fundamental physics with the space based GW detectors include testing GR in the ultra-strong regime, proving the existence of BH horizons, testing no-hair theorems and cosmic censorship, searching for scalar gravitational fields and other deviations from GR, looking for cosmic GW background, testing the order of the electroweak phase transitions and searching for cosmic strings. The low-frequency space detectors are sensitive to the frequency range 100 nHz-100 mHz and have high S/N ratios in detecting these sources. Schutz summarized the sensitivities of these GW detectors in a single slide (Fig. 2). From Fig. 2, we can see that the GWs from $10^5$ $M_{Sun}$ mergers and $10^6$ $M_{Sun}$ mergers have high S/N ratios for NGO/eLISA, LISA and ASTROD-GW detectors.

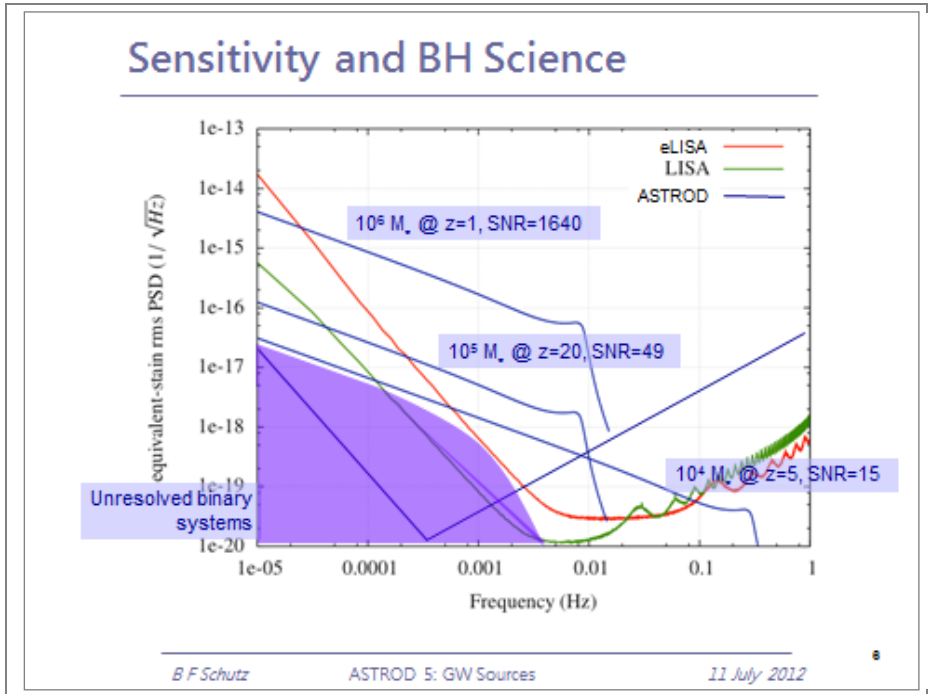

Fig. 2. Bernard Schutz's slide on sensitivity and BH science.

NGO/eLISA (www.elisa-ngo.org) with its 1 million km arm length is more sensitive in the higher frequency part of this range. It is 5 times down-scaled from LISA after the withdrawal of NASA from the project. Its thorough assessment study was presented by Oliver Jennrich from ESA. Paul McNamara reported the present status of LISA Pathfinder (LPF). This technological demonstrator LPF has tested most of its systems already and is scheduled for launch in the last quarter of 2014. With the outstanding review of ESA's Science Advisory Board and with the success of LPF, NGO/eLISA is expected to be launched around 2025-6.

ASTROD-GW (ASTROD [Astrodynamical Space Test of Relativity using Optical Devices] optimized for GW detection) is an optimization of ASTROD to focus on the goal of detection of GWs. The scientific aim is focused on GW detection at low frequency. The mission orbits of the 3 spacecraft forming a nearly equilateral triangular array are chosen to be near the Sun-Earth Lagrange points L3, L4 and L5 (Figure 3). The 3 spacecraft range interferometrically with one another with arm length about 260 million kilometers.

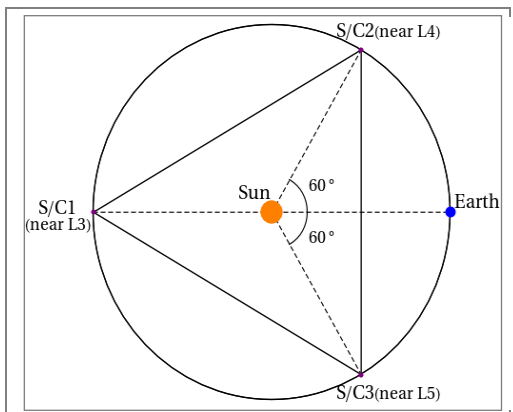

Fig. 3. Schematic of ASTROD-GW mission orbit configuration.

Since the sensitivities of the space interferometers are limited by local accelerometer noises in the low-frequency limit, the strain (δL/L) sensitivities are inversely proportional to the arm length. In Fig. 2, same local accelerometer noises are assumed. Due to arm length ratio, ASTROD-GW is better in sensitivity in the low-frequency limit by this ratio of 260 compared with NGO/eLISA. ASTROD-GW has the best sensitivity in the frequency band 100 nHz - 1 mHz. The weak light phase locking requirement due to longer distance is demonstrated in the laboratory optical activities of JPL. Fig. 2 shows a large part of the ASTROD-GW sensitive region is covered by unresolved binary confusion limit. Nevertheless, the larger S/N ratios for the more massive SMBH mergers facilitate luminosity distance determination and the association of electromagnetic counterparts for redshift measurements. This will enable a more precise determination of the equation of state for dark energy.

In his talk on ASTROD-GW, Ni from Tsing Hua University and Shanghai Normal University showed a diagram from the white paper of Demorest *et al.* (arXiv: 0902.2968) with the sensitivities of ASTROD-GW and inflationary GW (with tensor index n = 1, $\Omega_{gw} = 10^{-16}$) lay on it (Fig. 4). This diagram extrapolates Figure 2 to lower frequency and shows clearly the potential detectability of MBH-MBH background for ASTROD-GW. The detection of primordial (inflationary) GWs requires six-S/C configuration and this has a potential to probe three orders deeper into the MBH-MBH background (or foreground in this case; the WD binary confusion limit is dominated by this background near 100 nHz). PTA observations will give the level and characteristics of this background and one will be able to analyze the sensitivity of the six-S/C configuration to the primordial GWs. An-Ming Wu from NSPO (National Space Organization) present an analysis of deployment of the 3 ASTROD-GW S/C to their science destination. This is crucial for cost estimation of the mission.

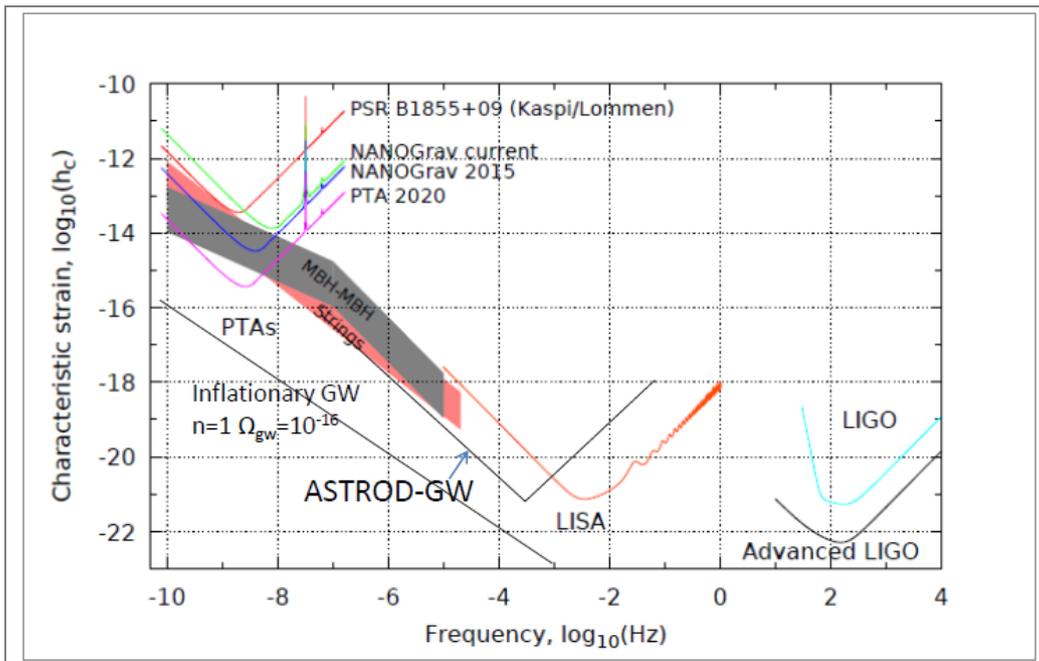

Fig. 4. Comparison of current and planned GW detectors, showing characteristic strain ($h_c$) sensitivity versus frequency along with expected source strengths (Demorest et al., 2009, arXiv: 0902.2968). LIGO, LISA and PTAs occupy complementary parts of the GW spectrum. *There is an outstanding gap in the detection band 100 nHz to 10 μHz.* The gray strip is the region all current models of MBH-MBH GW backgrounds occupy. ASTROD-GW has the best sensitivity in the 100 nHz – 1 mHz band. *The outstanding gap in the detection band 100 nHz to 10 µHz is filled by ASTROD-GW. The MBH-MBH GW backgrounds of all current models are above the ASTROD-GW sensitivity level.* The line in the bottom left corner corresponds to $\Omega_{gw} = 10^{-16}$ inflationary GW background ($\Omega_{gw}(f)$ in the figure is decadal density in terms of critical density of the universe defined to be = $(1/\rho c)$ $(d\rho_{gw}/d\log f)$ with $\rho_{gw}$ the energy density of the stochastic GW background and $\rho_c$ the present value of the critical density for closing the universe in general relativity.)

Masaki Ando from NAOJ described the DECI-hertz interferometer Gravitational wave Observatory (DECIGO) project in Japan. DECIGO will have a good sensitivity at around 0.1-10 Hz band with best sensitivity at 0.1 Hz, bridging the observation bands between ground-based GW detectors and eLISA. Its original and ultimate scientific goal is to detect primordial GWs. DECIGO is formed by three drag-free spacecraft, separated by 1,000 km from each other. It forms three Fabry-Perot interferometers with 1 m 100 kg mirrors and finesse 10. The laser (532 nm) power is 10 W. The arm length should be maintained to 20 pm. The three stages of DECIGO are DECIGO Pathfinder, pre-DECIGO and (full) DECIGO. The design and the current status of its milestone mission DECIGO Pathfinder mission were presented. Kent Yagi (MSU) gave a presentation on the science case for DECIGO Pathfinder and pre-DECIGO. The most interesting aspect being the direct observations of galactic intermediate black hole binaries whose masses and spins may be estimated with very good accuracy.

K G Arun from CMI gave an overview of various ideas which exist in the literature to use gravitational wave observations to perform *strong-field tests of General Relativity*. The focus was on the tests that are possible using the observations of inspiralling compact binaries with the ground-based and space-based GW interferometers. The important role of space-based GW detectors in performing precision tests of GR was underlined. Atish Kamble emphasized the astrophysical and cosmological importance of observing a *electromagnetic counterpart to a GW event associated with compact binary merger*. The chances of detect EM counterparts associated with compact binary mergers at radio wavelengths was discussed. Advent of several modern radio-band facilities and surveys coming up in the next few years can complement the GW observations of second generation interferometers (AdLIGO, AdVIRGO and KAGRA). Zakharov from ITEP reviewed the effects of gravitational lensing on GW propagation and detection. The effects would be more important for higher S/N observations and limit the precision of observations at high redshifts. More study is crucial for extraction of GW data and interpretation of GW observations.

Rajesh Nayak presented an overview of the algebraic formalism of *Time Delay Interferometry (TDI) for unequal-arm space based gravitational missions.* He especially presented cases with four beams. Ni talked on using relativistic ephemeris to numerically simulate the TDIs for ASTROD-GW and NGO/eLISA, and compare the simulated results of second-generation TDIs together with those of LISA and arm-length doubled NGO/eLISA. All the results satisfy their respective requirements.

ASTROD I is a planned interplanetary space mission with multiple goals. The primary aims are: to test general relativity with an improvement in sensitivity of over three orders of magnitude, improving our understanding of gravity and aiding the development of a new quantum gravity theory; to measure key solar system parameters with increased accuracy, advancing solar physics and our knowledge of the solar system; and to measure the time rate of change of the gravitational constant with an order of magnitude improvement and probing dark matter and dark energy gravitationally. Hanns Selig of ZARM, Bremen presented an overview of the mission concept, experimental setup, thermal aspects and an update of the mission study with a design of the spacecraft (Fig. 5). The mission concept is to have a drag-free spacecraft launched directly to solar orbit from low earth orbit with Venus gravity assistance during two encounters to reach the other side of the Sun (relative to Earth) in about a year. The ASTROD I S/C (spacecraft) will consist of one spacecraft carrying a telescope, four lasers, two event timers and a clock. Two-way, two wavelength laser pulse ranging will be used between the spacecraft in a solar orbit and deep space laser stations on Earth, to achieve the ASTROD I goals. In 2011, ASTROD I has been selected as one of the final 14 candidates for the Cosmic Vision M3 mission. It is an international project, and is envisaged as the first in a series of ASTROD missions. Finally the mission was not selected for the final 4 candidates for CV M3 Assessment Study in 2011. Nevertheless, ASTROD is a very promising concept for a fundamental physics space mission and shares some key technologies with other popular space missions like LPF/NGO and Jason 2 (T2L2). T2L2 on Jason 2 has been successful. The success of LPF will verify the crucial drag-free technology needed for ASTROD missions.

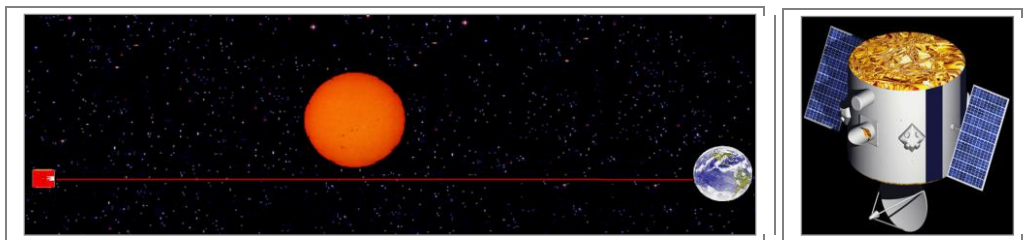

Fig. 5 ASTROD I orbit configuration and S/C drawing

Ions with energies larger than 100 MeV/n would charge test masses and induce spurious forces in drag-free missions. Discharge is necessary to maintain drag-free operations. Catia Grimani addressed to this issue and the placement and design of radiation monitors on the LISA Pathfinder (LISA-PF) mission. She also proposed to place similar devices on board the eLISA/NGO and ASTROD missions as well. These detectors allow us to monitor the integral flux of energetic particles penetrating mission spacecraft and inertial sensors at any time. She illustrated advantages and limits of these devices for estimating the actual test-mass charging. In addition, particle detectors on board space interferometers can be used to carry out multipoint observations of solar energetic particles (SEPs) at small and large solar longitudes at different distances from Earth with minor normalization errors. These measurements will provide important clues on solar physics and space weather investigations.

Clocks have been most precise measuring instruments. In this Symposium, Philippe Laurent of Observatoire de Paris talked on Cold Atoms devices to probe the Space Time with high accuracy: ACES space mission and STE-QUEST. The ACES (Atomic Clock Ensemble in Space) mission has first to demonstrate the high performances of a new generation of space clocks --- a cold cesium clock, called PHARAO, and an active hydrogen maser to be installed on an external pallet of the international space station (ISS). With high performance time transfer links to implement time and frequency comparisons with Earth based clocks with a time stability of 10 ps over ten days and a frequency accuracy of $10^{-16}$, the main objectives are testing the validity of the general relativity with accurate measurements of the gravitational red-shift and tests of the stability of the fundamental constants by comparing various ground based clocks. This mission is scheduled to operate from 2015 on. Based on an improved atomic clock, as compared to ACES, and an atomic interferometer able to test the equivalence principle by comparing different test masses, a future mission proposal STE-QUEST (Space Time Explorer and Quantum Equivalence Principle Space Test) is currently in the assessment phase in the frame of an M-class mission of the cosmic vision plan.

Over the past two decades measurements of temperature and polarization fluctuations in Cosmic microwave background (CMB) have spearheaded the impressive progress in cosmology. Tarun Souradeep of IUCAA reviewed the current status of the field dominated by results from the WMAP CMB mission. He then described the Planck Surveyor satellite (launched in 2010) and the many results it is expected to deliver in cosmology. The measurement/constraint on the ratio of tensor to scalar perturbations will indicate the magnitude of primordial tensor GWs which will give clues of detectability to ASTROD-GW and DECIGO missions.

Biswajit Paul (RRI) gave an overview of the science potential of the multi-wavelength satellite *ASTROSAT*. This satellite will be capable of doing high precision X ray timing in the 2-80 keV band with moderate energy resolution. The possibility of ASTROSAT complementing the GW experiments was also mentioned. A S Kirankumar (Space Applicationss Centre) gave a brief account of *the payload capabilities of Indian space missions*. Starting with ARYABHATTA, the first Indian satellite, ISRO has made a lot of progresses in its payload capabilities. India's recent Lunar mission Chandrayaan-I and upcoming mission Chandrayaan-II were also discussed.

Sreekumar chaired *a panel discussion on space-based GW detection*. LISA, DECIGO and ASTROD-GW were the three configurations that were considered. The planned payloads and the frequency ranges of observations of these three missions were compared. LISA Pathfinder is scheduled for launching in the last quarter of 2014; NGO/eLISA hopefully in 2025-6. DECIGO Pathfinder will be bidding for next JAXA selection in about 2-3 years. ASTROD-GW was proposed in 2009 upon call for proposals of GW mission studies by Chinese Academy of Sciences and is still looking for patrons. There was unanimous agreement that the success of LISA Pathfinder (which will test some of the crucial technologies all of these detectors will use) and the first detection of GWs by ground-based detectors would really improve the funding situation for these missions. There was a general feeling that there was need for international collaboration in order to achieve the final goals of space based gravitational wave detection.

G Srinivasan gave the concluding remarks. He saw LIGO-India as a window of opportunity for India to enter into contemporary big science. It could attract young people into science and engineering. He also felt that it was important that IndIGO joined the eLISA consortium. He said that this project gave a rare opportunity to cultivate a symbiosis between theorists, experimentalists and engineers in India. It poses challenges to the Indian industry and requires a cultural revolution. The most interesting aspect of GW detection with space based detectors, to him, was the power of these observations to test GR in the very strong field regime inaccessible till now.

He found the prospects of Pulsar Timing Arrays to detect GWs to be very interesting and not foreseen during the initial stages by the Pulsar community. The role of quantum clocks and coherent atoms in detecting GWs in space were also exciting. He suggested setting up a working group under the joint auspices of IAU, COSPAR and science academies of various countries interested in GW detection in space.

Invited talks of the ASTROD 5 Symposium will be published in a special issue (January, 2013) of review articles in IJMPD. The ASTROD5 meeting was hosted and financially supported by the Raman Research Institute, Bangalore. Outstation Student participation was facilitated by travel grants from the ASTROD Foundation.

K.G. Arun, Chennai Mathematical Institute, Chennai, India
Bala R. Iyer, Raman Research Institute, Bangalore, India
Wei-Tou Ni, National Tsing Hua University, Hsinchu, Taiwan, ROC

1 August 2012